\documentstyle [preprint,aps]{revtex}
\begin{document}
\draft

\title
{Correspondence in Quasiperiodic and Chaotic Maps: \\
Quantization via the von Neumann Equation}
\author{Joshua Wilkie and Paul Brumer}
\address{Chemical Physics Theory Group, Department of Chemistry,
   University of Toronto,
   Toronto, Ontario, Canada M5S 1A1 }

\maketitle

\begin{abstract} A generalized approach to the quantization of a large
class of maps on a torus, i.e. quantization via the von Neumann Equation,
is described and a number of issues related to the quantization of model
systems are discussed. The approach yields well behaved mixed quantum
states for tori for which the corresponding Schrodinger equation has no
solutions, as well as an extended spectrum for tori where the Schrodinger
equation can be solved. Quantum-classical correspondence is demonstrated
for the class of mappings considered, with the Wigner-Weyl density
$\rho(p,q,t)$ going to the correct classical limit.  An application to the
cat map yields, in a direct manner, nonchaotic quantum dynamics, plus the
exact chaotic classical propagator in the correspondence limit.

\end{abstract}


\section{Introduction}

Quantization of systems with classical analogs proceed by a well known
procedure. Specifically, one replaces coordinates or momenta by operators
in the Hamiltonian, constructs the Schrodinger equation and imposes
boundary conditions which arise directly from the classical physics.  In
this paper we show for the case of maps on a torus, model systems in
nonlinear dynamics, that modifications to this procedure are essential to
allow the quantum dynamics to unambiguously approach the correct classical
limit.

Area preserving mappings on the torus are model systems which have long
been used to illustrate the essence of regular or chaotic classical
dynamics because they are of low dimensionality and can be readily
propagated. These maps are essentially systems which are confined to finite
coordinate and momentum space regions $0 \le q < a, 0 \le p < b$, with
rules of dynamic evolution which confine the system to this region through
a dependence on $q$ mod $a$, $p$ mod $b$. In addition, time evolution is
in discrete steps. Much of the current interest\cite{ford}-\cite{balazs2} in
quantizing these mappings stems from the expectation that they will prove
useful in understanding the role of classical chaos in quantum dynamics
and in understanding quantum-classical correspondence. Indeed, little is
known about quantum-classical correspondence for systems whose classical
analog is chaotic\cite{berry2}. While the correspondence principle requires
that
quantum mechanics should produce the classical laws in the limit that
Planck's constant approaches zero\cite{cp}, the characteristics of the
limit are such as to make this verification extremely difficult\cite{berry2}.

Central to the issues of classical-quantum correspondence is the specific
prescription used to quantize a given classical system.  In this paper we
show that a new extended quantization procedure allows an alternative
method of quantizing mappings on a torus which yields solutions having a
number of desirable features which do not arise in the traditional
quantization via the Schrodinger equation.

Motivation for reconsidering the quantization procedure for maps on a torus
stems from an examination of the results of previous efforts in this area.
Hannay and Berry\cite{berry}, and Balazs and Voros\cite{balazs}, chose to
directly quantize the Schrodinger propagator while Ford and
coworkers\cite{ford} quantize certain maps by introducing a kicked
oscillator Hamiltonian ($-\infty <q,p<\infty$)
\begin{equation}
H=p^2/2\mu +\epsilon q^2/2\sum_{s=-\infty}^{\infty}\delta(s-t/T),
\label{fh}
\end{equation}
The dynamics generated by this Hamiltonian stroboscopically produces the
classical dynamics of the map. They
then construct the propagator for this system.  The restriction of the
classical dynamics to a torus $0\leq q < a$, $0\leq p < b$ is introduced
by imposing periodic boundary conditions on the wavefunctions in both the
position and momentum representations. The results of this procedure have
been the subject of considerable controversy. Specifically,
Ford\cite{ford} showed that although classical cat map dynamics is
algorithmically complex, the quantum cat does not show this type of random
behavior in the classical limit.  Berry and coworkers, however,
argue\cite{berry2} that such measures are less useful indicators of
correspondence failure than believed by Ford and coworkers. Further,
Graffi et al\cite{graffi} have recently shown that the quantum cat is
mixing in a specific limit, $N \rightarrow \infty$ along the subsequence
of $N$ prime, where $h = 1/N$. This is in contrast with Ford's
demonstration of quasiperiodic dynamics when taking a somewhat different
approach to the classical limit.

Clearly the traditional quantization procedure leads to a quantum cat map
whose behavior as $h \rightarrow 0$ is, at best, subtle. In addition the
traditional approach leads to two other features which we regard as
undesirable:  (a) only a restricted class of classical tori, those with
$ab=h N$, $N$ integer, can be quantized\cite{models}. Indeed, some
mappings can only be quantized for $N$ even\cite{ford,balazs}; and (b) the
resultant wavefunctions are highly singular, being a set of delta
functions.

In this paper we show that introducing a translationally invariant
Hamiltonian, quantizing the associated von Neumann (quantum Liouville)
equation $ \partial \hat{\rho}/\partial t = [H,\hat{\rho}]/i\hbar $
for the density matrix $\hat{\rho}$,
and obtaining solutions satisfying Bloch boundary conditions,
allows quantization of tori for all $ab$, is physically complete, leads to
smooth well behaved solutions and is measure preserving.  Further, we show
that the correct classical propagator and the exact classical dynamics
emerge directly from the quantum propagator in
the classical limit, with the limit taken as the complete sequence $N
\rightarrow \infty$.
Hence the classical limit of our quantum propagator for the cat map
generates algorithmically complex dynamics, the system is mixing, etc.

Note that our approach yields density matrix solutions
for cases (e.g., $ab \ne hN$)  for which there are
no pure states. The fact that
density matrix solutions exist when wavefunctions do not is of considerable
interest and is emphasized below.

This paper is organized as follows: Section II summarizes aspects of the
traditional quantization procedure and treats the case of $\epsilon=0$ in
Eq. (1) explicitly. The extended quantization procedure is then introduced
in Section III where it is also applied to the case of Eq. (1) with
$\epsilon=0$. Examining this case allows us to emphasize the different
spectra and solutions obtained by this approach and to show that both
eigenfunctions and eigenvalues of the von Neumann equation directly
approach the classical limit.  In Section IV  we generalize the discussion
to the case of $\epsilon\neq 0$ and use the von Neumann quantization
scheme to obtain the quantum propagator. Classical and quantum map
dynamics are expressed in a form suitable for the study of correspondence.
Section V then explores the quantum dynamics in the correspondence limit
where we show that all maps treated reduce to the correct classical limit. In
the application to the quantum cat we show that its quantum dynamics is
not ergodic, but that the chaotic classical cat is nonetheless recovered
in the $h\rightarrow 0$ limit. Throughout this treatment we make extensive
use of the Wigner-Weyl representation \cite{ww,jaffe}, a $p,q$ dependent
representation of quantum mechanics whose form encourages a direct
comparison with classical dynamics. In Section VI we address the
question of why there are no pure states for the quantum cat, showing that
the system best resembles a discretized Langevin process. Finally, Section
VII contains a summary and remarks on future work.

\section{Traditional Quantization}

Constructing the appropriate quantum generalization of a given classical
system requires a correspondence rule for replacing classical observables
by quantum operators, a choice of a dynamical state equation and a
statement of the associated boundary conditions.  The appropriate
dynamical state equations are normally assumed to be the Schrodinger
equation for pure state dynamics and the von Neumann equation
(or Quantum Liouville Equation) for mixed
state dynamics and, customarily, the choice of boundary condition is
dictated by features of the physical system.  Quantum mappings on the
torus have no known physical realization, and so some freedom would appear
to exist for the choice of boundary conditions. Below we discuss the
limitations of quantization via the Schrodinger equation, both for
(previously adopted) periodic boundary conditions
as well as for Bloch boundary conditions.
We later show that a physically complete theory
of quantum mappings on a torus does result from a von Neumann based
quantization procedure in concert with Bloch type boundary conditions.
Of general interest is the different way in which the boundary conditions,
implemented in these two procedures and both apparently consistent with
the classical map, affect the quantum picture.

\subsection{Boundary Conditions}

Traditional quantization procedures\cite{ford,berry} define the dynamics
on the full phase
space, $-\infty < p,q < \infty $, but impose periodic boundary conditions
on the wavefunction
in both the coordinate $[ \psi(q) ]$ and momentum representations
$ [ \overline{\psi}(p) ]$, i.e.
\begin{equation}
\psi(q) = \psi(q+a) \equiv T_q(a) \psi(q) \label{bc1}
\end{equation}
\begin{equation}
\overline{\psi}(p) = \overline{\psi}(p+b) \equiv T_p(b) \overline{\psi}(p)
\label{bc2}
\end{equation}
where  Eqs. (\ref{bc1}) and (\ref{bc2}) implicitly define the coordinate and
momentum
translation operators
\begin{equation}
T_q(a)= \exp (i\hat{p}a/\hbar) ;\ T_p(b) = \exp(-i\hat{q}b/\hbar).
\end{equation}
As a consequence of Eqs. (\ref{bc1}) and (\ref{bc2}),
dynamics within each $(q,p)$ unit cell is expected to reflect the
character of torus dynamics [i.e. dynamics with variables (q mod a)
and (p mod b)]. From Eq. (\ref{bc2}) it follows that
$\{1-e^{-ibq/\hbar}\}\psi(q)=0$
which implies that $\psi(q)=0$ unless $q=nh/b$ where $n\in
\bf{Z}$. Similarly, from Eq. (\ref{bc1}) it follows that
$\{1-e^{iap/\hbar}\}\bar{\psi}(p)=0$
which implies $\bar{\psi}(p)=0$ unless $p=mh/a$ where
$m\in\bf{Z}$. Imposing both conditions simultaneously gives
$\{1-e^{iab/\hbar}\}\psi(q)=0$ and $\{1-e^{-iab/\hbar}\}\bar{\psi}(p)=0$
so that we must have $ab=h N$ where $N\in\bf{Z}$. Fixing $N$
one then obtains the class of allowed wavefunctions
\begin{equation}
\psi(q)=\sqrt{a}\sum_{j=-\infty}^{\infty}\psi_j\delta(q-ja/N)
\label{waveq}
\end{equation}
with $\psi_{j+N}=\psi_j$. This condition implies that
\begin{equation}
\psi_j=\sum_{l=1}^N\alpha_le^{2\pi ijl/N},
\end{equation}
where the $\alpha_l$ coefficients are arbitrary. In the
momentum representation these states take the form
\begin{equation}
\bar{\psi}(p)=\sqrt{b}\sum_{j=-\infty}^{\infty}\sum_{k=1}^N\bar{\psi}_k\delta(p-(j+k/N)b)
\label{wavep}
\end{equation}
where $\bar{\psi}_k=\sqrt{N}\alpha_k$.
Thus, in general (independent of the
Hamiltonian $H$) periodic boundary conditions in $p,q$ lead to a discrete
spectrum with
non-normalizable wavefunctions, comprised of sets of delta functions.
The traditional interpretation of
$|\psi|^2$ as a probability is inapplicable.
Note that it is impossible to quantize any mapping on a torus by this
route unless $ab = hN$, a consequence of the Fourier transform relationship
between $p$ and $q$ and of the requirement to simultaneously fit waves in both
the $q$ and $p$ directions. With $ab \ne hN$ neither the Hamiltonian, nor
any other operator, has eigenvalues. The quantization procedure is thus
incomplete insofar as only a small number of classical tori can be
quantized\cite{models}.	Finally, note that these conclusions are
independent of the system and so apply to any mapping confined to a torus.

The Wigner-Weyl picture \cite{ww,jaffe} provides a $p,q$ dependent quantum
representation in the density matrix formulation which allows a direct
comparison with classical mechanics in phase space.  One can readily
demonstrate that in this picture the dynamics lies on points in phase
space which are rational multiples of $a$ and $b$. Ford shows\cite{ford}
that this behavior survives in the classical limit and is
responsible for non-algorithmically complex dynamics for the classical
limit of the quantum cat.  It is obviously also the case for any map
quantized by this procedure.

The situation is similar if Bloch type boundary conditions are adopted.
That is, Bloch boundary conditions on the wavefunction
\begin{eqnarray}
\psi(q+a)&=&e^{ip_0a/\hbar}\psi(q) \nonumber \\
\bar{\psi}(p+b)&=&e^{-iq_0b/\hbar}\bar{\psi}(p)
\label{Blochbc}
\end{eqnarray}
with constant $q_0,p_0$
again yield states
\begin{equation}
\psi(q)=\sqrt{a}\sum_{j=-\infty}^{\infty}\psi_j\delta(q-q_0-ja/N)
\end{equation}
only when $ab=hN$. Here
$\psi_j=e^{ip_0aj/N\hbar}\sum_{l=1}^N\alpha_le^{2\pi ijl/N}$ where the
$\alpha_l$ coefficients are arbitrary. In the momentum representation
\begin{equation}
\bar{\psi}(p)=\sqrt{b}e^{-ipq_0/\hbar}\sum_{j=-\infty}^{\infty}
\sum_{k=1}^N\bar{\psi}_k\delta(p-p_0-(j+k/N)b)
\end{equation}
where $\bar{\psi}_k=\sqrt{N}\alpha_k$. Bloch boundary conditions will,
however, prove useful in the extended quantization scheme discussed below.

\subsection{The case of $\epsilon=0$}

Consider now, for illustration purposes, the case of $\epsilon=0$. Here
Eq. (1) gives  $N=ab/h$ independent Hamiltonian eigenfunctions\cite{foot1}$^a$

\begin{equation}
\overline{\psi}_l(p) = \sqrt{b/N} \sum_{j=-\infty}^{\infty} \delta(p-(j+l/N)b)
\label{e0eigf}
\end{equation}
and associated discrete spectrum $E_l = (lb/N)^2/2\mu$, $l=1,...,N$.
We also consider, for comparison with results below, the solution from the
perspective of the von Neumann (or quantum Liouville) equation:
\begin{equation}
\partial \hat{\rho}/\partial t = -i [H,\hat{\rho}]/\hbar \equiv -i L_q
\hat{\rho}.
\label{vonN}
\end{equation}
Here $\hat{\rho}$ is the density operator and Eq. (\ref{vonN}) defines the
quantum Liouville operator $L_q$. The eigenvalues $\lambda_{k,l}$ and
eigenfunctions $\hat{\rho}_{k,l}$ of the $L_q$ satisfy:
\begin{equation}
L_q \hat{\rho}_{k,l}= \frac{1}{\hbar}[H,\hat{\rho}_{k,l}]  =
\lambda_{k,l} \hat{\rho}_{k,l}
\end{equation}
Both may be obtained directly from the
Hamiltonian solutions and the latter conveniently
expressed in the Wigner-Weyl representation.
Specifically,  $(k,l =1,...,N)$:
\begin{equation}
\lambda_{k,l} = (E_k - E_l)/\hbar = (k^2 -l^2)b^2/(2 \mu N^2)
\label{wweigv}
\end{equation}
and $\hat{\rho}_{k,l}= |\psi_k><\psi_l|$ so that the Wigner-Weyl
representation of $\hat{\rho}_{k,l}$, denoted $\rho_{k,l}(p,q)$, is given
by \cite{ww,jaffe}
\begin{eqnarray}
\rho_{k,l}(p,q)&=& h^{-1} \int_{-\infty}^{\infty} dq'
<q-q'/2|\psi_k><\psi_l|q+q'/2>e^{ipq'/\hbar} \nonumber \\
&=&h^{-1} \int_{-\infty}^{\infty} dp'
<p+p'/2|\psi_k><\psi_l|p-p'/2>e^{iqp'/\hbar}.
\label{wweq1}
\end{eqnarray}
Using Eq. (\ref{e0eigf}) we have
\begin{equation}
\rho_{k,l}(p,q) = 2a^{-1} \sum_{n,m=-\infty}^{\infty} e^{2\pi i(n-m)qN/a}
e^{2\pi i(k-l)q/a} \delta(2p-(n+m+\frac{k+l}{N})b).
\label{eigfeigv}
\end{equation}
A simple change in the integers of summation in this expression for
$\rho_{k,l}(p,q)$ and use of the identity\cite{vlad}
\begin{equation}
\sum_{j=-\infty}^{\infty}e^{2\pi ijx}=\sum_{j=-\infty}^{\infty}\delta(x-j)
\label{id1}
\end{equation}
gives
\begin{equation}
\rho_{k,l}(p,q)= (2N)^{-1} e^{2\pi i(k-l)q/a} \sum_{n,m=-\infty}^{\infty}
(-1)^{mn}\delta(q-ma/2N)\delta(p-(n+\frac{k+l}{N})b/2).
\end{equation}
Thus, the eigenfunctions of $L_q$ in this representation are delta functions
localized on a discrete number of points of the torus which are
rational multiples of $a$ and $b$, i.e., a delta function ``brush" in
phase space. Note that
Eq. (\ref{id1}) is repeatedly used throughout this paper although it is
not always cited.

The results in this subsection are used later below for comparison with
the extended quantization results.

\section{Extended Quantization}

An examination of a different approach to map quantization procedure is
hence well motivated.
The extended quantization procedure which we advocate entails three new
features: (a) the introduction of ``mod operators" which produce a properly
symmetrized Hamiltonian, (b) solving the von Neumann [Eq. (\ref{vonN})], rather
than
Schrodinger, equation and (c) use of Bloch boundary conditions on the
density matrix. The latter is motivated immediately below. [Throughout
this paper, as above, we denote quantum density operators by
$\hat{\rho}$ and the Wigner-Weyl representation by $\rho(p,q)$. In addition,
analog quantities in classical mechanics, such as the phase space
density, will carry a ``c" superscript.]

\subsection{Boundary conditions}

Consider then the von Neumann equation with either of
the two choices of boundary condition i.e. periodic and Bloch type.
 From the perspective of the density matrix, periodic boundary conditions
on the wavefunction imply
$T \hat{\rho}= \hat{\rho}, \hat{\rho} T = \hat{\rho}$ for $T=T_q(a)$ and
$T=T_p(b)$.
In the Wigner-Weyl representation\cite{ww}
\begin{equation}
\rho(p,q) = h^{-1} \int_{-\infty}^{\infty} dq'
<q-q'/2|\hat{\rho}|q+q'/2>e^{ipq'/\hbar}
\end{equation}
these conditions take the form
\begin{eqnarray}
e^{ipa/\hbar}\rho(p,q\pm a/2) &=& e^{-ipa/\hbar}\rho(p,q\pm a/2) =\rho(p,q)
\nonumber \\
e^{iqb/\hbar}\rho(p\pm b/2,q) &=& e^{-iqb/\hbar}\rho(p\pm b/2,q)=\rho(p,q).
\label{perbc}
\end{eqnarray}

The following requirements on the density result from Eq. (\ref{perbc}): (a)
$(1-e^{i2qb/\hbar})\rho(p,q)=0$ which implies
$q=hl/2b$ where $l\in {\bf Z}$, (b) $(1-e^{i2pa/\hbar})\rho(p,q)=0$
which implies $p=hk/2a$ where $k\in{\bf Z}$, and (c)
$(1-e^{iab/\hbar})\rho(p,q)=0$ which implies that $ab=hN$ where $N\in
{\bf Z}$. The resulting states are mixtures of the delta type
wavefunctions of Eq. (\ref{waveq}) localized on the rational points of
the torus and they exist only when $ab=hN$. Hence this combination of
dynamic law and boundary condition regenerate the same difficulty as that
associated with the Schrodinger approach and is, once again, independent of
the system. This is the case since Eq. (\ref{perbc}) is an exact statement
of the requirement of periodicity on the wavefunction, transferred to the
density matrix. It results in precisely the same restriction as it did in
its application in the wavefunction picture\cite{strangebc}.

A similar situation does not, however, arise with the application of Bloch
boundary conditions to the density matrix $\rho$. Specifically, here
Eq. (\ref{Blochbc}) implies that
\begin{equation}
<q+a|\hat{\rho}_{jk}|q+a> = <q+a|\psi_j><\psi_k|q+a> =
<q|\psi_j><\psi_k|q>=<q|\hat{\rho}_{jk}|q>
\end{equation}
with a similar result in the momentum space picture. Thus, the phase
$\exp(ip_0a/\hbar)$ associated with the wavefunction drops out, resulting
in a quite different boundary condition in the density matrix picture than
in the wavefunction picture. Specifically, the conditions
$T^{-1}\hat{\rho}T=\hat{\rho}$ imply Bloch boundary conditions on
the wavefunction $\psi$ if $\hat{\rho}=|\psi\rangle\langle\psi |$, but
densities $\hat{\rho}$ can exist which satisfy
$T^{-1}\hat{\rho}T=\hat{\rho}$ and
which do not take the form of a weighted sum over pure states satisfying
Bloch boundary conditions on the wavefunction. These boundary conditions
will prove extremely
useful in conjunction with the introduction of mod operators, as described
below.

\subsection{Mod Operators}

We consider solutions to the
von Neumann equation with Bloch type boundary conditions.
To do so we first introduce the operators
($\widehat{p{\rm mod}~b}$) and ($\widehat{q{\rm mod}~a}$). These are
obtained as a direct extension of the classical Fourier expansions for ($q$
mod $a$) and ($p$ mod $b$), i.e. as extensions of :
\begin{eqnarray}
(q{\rm mod}~a)/a &=&
Q{\rm mod}~1=\frac{1}{2}+\sum_{m=-\infty\atop m\neq 0}^{\
\infty}\frac{i}{2\pi m}f_{0,m} \nonumber \\
(p{\rm mod}~b)/b &=&
P{\rm mod}~1=\frac{1}{2}+\sum_{n=-\infty\atop n\neq 0}^{\
\infty}\frac{i}{2\pi n}f_{n,0},
\label{class}
\end{eqnarray}
where $f_{n,m}(P,Q)=\exp[2\pi i(nP+mQ)]$ and where we have
introduced the convenient dimensionless variables
$Q=q/a, P=p/b$. Specifically, the quantum operators, analogs of Eq.
(\ref{class}), are
\begin{equation}
(\widehat{p{\rm mod}~b})/b =
\widehat{P{\rm mod}~1}=\frac{1}{2}+\sum_{n=-\infty\atop n\neq 0}^{\
\infty}\frac{i}{2\pi n}\hat{f}_{n,0},
\label{modscl}
\end{equation}
with eigenvalues $P{\rm mod}~1$ and conjugate variable\cite{commrel}$^{a}$
\begin{equation}
(\widehat{q{\rm mod}~a})/a =
\widehat{Q{\rm mod}~1}=\frac{1}{2}+\sum_{m=-\infty\atop m\neq 0}^{\
\infty}\frac{i}{2\pi m}\hat{f}_{0,m}
\label{modsqm}
\end{equation}
with eigenvalues $Q{\rm mod}~1$. Here the unitary operators
$\hat{f}_{n,m}$ are obtained by Weyl quantizing\cite{ww}
$f_{n,m}(P,Q)=\exp[2\pi i(nP+mQ)]$, i.e. :
\begin{eqnarray}
\hat{f}_{n,m} &=& h^{-1} \int dp \int dq e^{2\pi i(nP+mQ)}
\int dv e^{ipv/\hbar} |q+v/2\rangle\langle q-v/2| \nonumber \\
&=&\int_{-\infty}^{\infty}dq~e^{2\pi
i mQ}|q-n\alpha a/2\rangle\langle q+n\alpha a/2| \nonumber \\
&=& \int_{-\infty}^{\infty}dp~e^{2\pi
i nP}|p+m\alpha b/2\rangle\langle p-m\alpha b/2|
\label{fnm}
\end{eqnarray}
where $\alpha=h/ab$ is a dimensionless form of Planck's constant.
Although  Eq. (\ref{fnm}) provides forms most useful for computations,
applying the coordinate translation operator inside the integral allows us
to rewrite Eq. (\ref{fnm}) in a far more attractive form:
\begin{equation}
\hat{f}_{n,m}= \exp\{2\pi i(n \hat{P}+m\hat{Q)}\}.
\label{qfb}
\end{equation}
where $(\hat{Q},\hat{P}) =(\hat{q}/a,\hat{p}/b)$ are the scaled coordinate
and momentum operators.
The $\hat{f}_{n,m}$ operators\cite{knabe}, whose properties are discussed in
Appendix
A, will allow us to write the quantum operator analog of the Fourier
transforms necessary to treat periodic systems on tori.

The introduction of these operators allows us to revise the Hamiltonian in
Eq. (1) to read:
\begin{equation}
H = (\widehat{p{\rm mod}~b})^2/2\mu + \epsilon (\widehat{q{\rm mod}~a})^2/2
\sum_{s=-\infty}^{\infty}\delta(s-t/T).
\label{newH}
\end{equation}
This new Hamiltonian satisfies $[H,T_p(b)]=[H,T_q(a)]=0$, i.e. $H$
reflects the desired
cell-like character of the full phase space. This character is also
demanded of the eigenfunctions $\hat{\rho}_{i,j}$ of the von Neumann equation,
i.e.
by requiring the generalized Bloch boundary conditions on $\hat{\rho}$,
\begin{eqnarray}
\hat{\rho}&=& T_p^{-1}(b) \hat{\rho} T_p(b) \nonumber \\
\hat{\rho}&=& T_q^{-1}(a) \hat{\rho} T_q(a).
\label{bc}
\end{eqnarray}
These conditions have the form, in the Wigner-Weyl representation, of:
\begin{eqnarray}
\rho(p,q+a) &=& \rho(p,q) \nonumber \\
\rho(p+b,q) &=& \rho(p,q).
\label{classbc}
\end{eqnarray}
Hence these conditions
correspond precisely to the boundary conditions one would
impose on a classical phase space density if the density (rather than the
wavefunction) were periodic in $p,q$.

States satisfying Eqs.(\ref{classbc}) must, by Fourier's Theorem, be of the
form
\begin{equation}
\rho(p,q)=\frac{1}{ab}\sum_{n,m=-\infty}^{\infty}\rho_{n,m}f_{n,m}=
\frac{1}{ab}\sum_{n,m=-\infty}^{\infty}\rho_{n,m}
\exp\{2\pi i(nP+mQ)\}
\label{form}
\end{equation}
in the Wigner-Weyl representation and so distributions satisfying
Bloch boundary conditions are expected to satisfy the analogous quantum Fourier
expansion:
\begin{equation}
\hat{\rho}=\frac{1}{ab}\sum_{n,m=-\infty}^{\infty}\rho_{n,m}\hat{f}_{n,m}
=\frac{1}{ab}\sum_{n,m=-\infty}^{\infty}\rho_{n,m}
\exp\{2\pi i(n \hat{P}+m\hat{Q)}\}.
\label{qfe}
\end{equation}
We utilize this approach below to examine the case of $\epsilon=0$.
Specifically, we solve the von Neumann equation  [Eq. (\ref{vonN})] using
the Hamiltonian of Eq. (\ref{newH}) coupled with the Bloch boundary
conditions which require Eq. (\ref{qfe}). The
general case $\epsilon \ne 0$ is considered in Section III.

\subsection{The case of $\epsilon=0$}

Consider the extended quantization procedure applied to the case
of $\epsilon=0$. The relevant Hamiltonian is then\cite{foot1}$^b$
\begin{equation}
H=(\widehat{p{\rm mod}~b})^2/2\mu = b^2 (\widehat{P{\rm mod}~1})^2/2\mu
\end{equation}
and the associated von Neumann equation
\begin{equation}
L_q \hat{\rho} = \frac{b^2}{2\mu\hbar}[(\widehat{P{\rm
mod}~1})^2,\hat{\rho}]
= i \partial \hat{\rho}/\partial t.
\label{vonNzero}
\end{equation}

Eigenfunctions of the Liouville Operator satisfy:
\begin{equation}
L_q\hat{\rho}=\lambda\hat{\rho}.
\label{qle}
\end{equation}
Inserting Eqs. (\ref{qfe}) and (\ref{vonNzero}) into Eq. (\ref{qle}) we obtain
\begin{equation}
\frac{b^2}{2\mu\hbar}\sum_{n,m=-\infty}^{\infty}\rho_{n,m}[(\widehat{P{\rm
mod}~1})^2,\hat{f}_{n,m}]=
\lambda\sum_{n,m=-\infty}^{\infty}\rho_{n,m}\hat{f}_{n,m}.
\end{equation}
Using Eq. (\ref{modscl}) it can be readily shown that, for arbitrary
$P_0$ (i.e. arbitrary $p_0/b)$,
\begin{eqnarray}
&&\sum_{n=-\infty}^{\infty}e^{-2\pi in P_0}
[(\widehat{P{\rm mod}~1})^2,\hat{f}_{n,m}]\nonumber \\
&=&\{[(P_0+m\alpha/2){\rm
mod}~1]^2-[(P_0-m\alpha/2){\rm
mod}~1]^2\}\sum_{n=-\infty}^{\infty}e^{-2\pi inP_0}\hat{f}_{n,m}.
\end{eqnarray}
Hence we can choose $\rho_{n,m}=\delta_{m,j}e^{-2\pi inP_0}$.
The eigenfunctions and eigenvalues of $L_q$ with Bloch boundary conditions
are then given by:
\begin{equation}
\hat{\rho}_{j,p_0}=(ab)^{-1} \sum_{n=-\infty}^{\infty}e^{2\pi
i[n(\hat{p}-p_0)/b
+j\hat{q}/a]}
\label{eq31}
\end{equation}
or, alternatively, using the momentum representation [Eq. (\ref{fnm})]
of $\hat{f}_{n,m}$ and  Eq. (\ref{id1})
\begin{equation}
\hat{\rho}_{j,p_0}=a^{-1}\sum_{n=-\infty}^{\infty}|p_0+nb+\pi
j\hbar/a\rangle\langle p_0+nb-\pi j\hbar/a|
\label{eigenf}
\end{equation}
with eigenvalues
\begin{equation}
\lambda_{j,p_0}=\frac{b^2}{2\mu\hbar}\{[(\frac{p_0}{b}+\frac{\pi
j\hbar}{ab}){\rm mod}~1]^2-[(\frac{p_0}{b}-\frac{\pi
j\hbar}{ab}){\rm mod}~1]^2\},
\label{eigene}
\end{equation}
where $j$ is an integer, $0 \le p_0 < b$,  and the arguments of the bras and
kets in Eq. (\ref{eigenf}) correspond to eigenvalues of $\hat{p}$.
The Wigner-Weyl representation $\rho_{j,p_0}(p,q)$ of these solutions
makes evident their character:
\begin{equation}
\rho_{j,p_0}(p,q) = a^{-1} e^{2\pi ijq/a} \sum_{n=-\infty}^{\infty}
\delta(p-p_0-nb).
\label{eigenfww}
\end{equation}

Stationary states ($\lambda_{j,p_0}=0)$ are of two types, arising from
either $j=0$ or $p_0=0$. The stationary states with $j=0$ are uniform in $q$
and belong to the point spectrum, whereas the $p,q$ integral over the
stationary states with $j\neq 0$ and $p_0=0$ are zero, characteristic of
elements of the
continuous spectrum. The overall spectrum of $L_q$ is continuous, due to the
continuous $p_0$ label.

Several aspects of these solutions are worth emphasizing. First,
Eqs. (\ref{eigenf}) and (\ref{eigene}) provide solutions for all values of
$a$ and $b$.  Second, the eigenfunctions in Eq. (\ref{eigenf}) are, in
general, mixed states, since wavefunctions exist only for $ab=hN$ whereas
Eq. (\ref{eigenf}) applies to all $ab$. If the
condition $ab=hN$ is satisfied then it is possible to recombine some of the
degenerate Liouville eigenstates to produce pure states. Even then,
however, the pure states are only a small fraction of all of the possible
solutions.

To permit a comparison of Eqs. (\ref{eigfeigv}), (\ref{eigenf}) and
(\ref{eigene}) to the classical limit,
consider solutions to the classical Liouville equation for motion on a
torus with the appropriate classical boundary conditions
[which are the same as Eq. (\ref{classbc})]. That
is\cite{jaffe} we consider the classical dynamics via the Liouville
equation:
\begin{equation}
\partial \rho^{(c)}/\partial t = \{H,\rho^{(c)}\} \equiv -iL_c \rho^{(c)}
\end{equation}
Here $\{,\}$ denotes the Poisson bracket, $\rho^{(c)}$ is the classical
phase space density and this equation defines the classical Liouville
operator $L_c$. Eigenfunctions and eigenvalues of $L_c$ are given by the
solutions to
the classical Liouville eigenvalue problem
\begin{equation}
L_c\rho^{(c)} =
-i(p{\rm mod}~b/\mu)\partial\rho^{(c)}/\partial q = \lambda^{(c)}
\rho^{(c)} .
\label{classL}
\end{equation}
Solutions to Eq. (\ref{classL}) are expected to satisfy Eq.
(\ref{classbc}). Hence $\rho^{(c)} $
must take the form of Eq. (\ref{form}) and so  (with classical Fourier
coefficients being denoted $\rho^{(c)}_{n,m})$
\begin{equation}
\sum_{n,m=-\infty}^{\infty}\rho^{(c)}_{n,m}[\frac{2\pi (p{\rm mod}~b)m}{\mu
a}-\lambda^{(c)}]e^{2\pi i(np/b+mq/a)}=0.
\end{equation}
Using Eq. (\ref{modscl}) and taking matrix elements in the Fourier basis
\begin{equation}
[\lambda^{(c)}-\frac{\pi mb}{\mu a}]\rho^{(c)}_{n,m}=-i\frac{mb}{\mu
a}\sum_{k=-\infty \atop k\neq n}^{\infty}\frac{\rho^{(c)}_{k,m}}{k-n}.
\label{eige}
\end{equation}
We assume a solution of the form $\rho_{n,m}^{(c)}=\Omega_m e^{-in\theta_m}$
and substitute this expression into Eq. (\ref{eige}). We obtain
\begin{equation}
[\lambda^{(c)}-\frac{\pi mb}{\mu a}]\Omega_me^{-in\theta_m}=-i\frac{mb}{\mu
a}\Omega_me^{-in\theta_m}\sum_{k=-\infty\atop k\neq
0}^{\infty}\frac{e^{-ik\theta_m}}{k}
\end{equation}
or
\begin{equation}
\lambda^{(c)}-\frac{\pi mb}{\mu a}=-\frac{2mb}{\mu
a}\sum_{k=1}^{\infty}\frac{\sin(k\theta_m)}{k}.
\end{equation}
Using the identity\cite{abra}
\begin{equation}
\sum_{k=1}^{\infty}\frac{\sin(k\theta)}{k}=\frac{\pi-\theta}{2}
\end{equation}
we have
\begin{equation}
\theta_m=\frac{\mu a\lambda^{(c)}}{mb}.
\end{equation}
Defining $\theta_m/2\pi = p_0/b$, $p_0\in[0,b)$, and choosing
$\Omega_m=\delta_{m,j}$ so that $\rho^{(c)}_{n,m}=\delta_{m,j}e^{-2\pi
inp_0/b}$ gives resultant classical eigensolutions
\begin{eqnarray}
\rho^{(c)}_{j,p_0}(p,q) &=& (ab)^{-1}e^{2\pi ijq/a}\sum_{n=-\infty}^{\infty}
e^{2\pi i n(p-p_0)/b} \nonumber \\
 &=& a^{-1} e^{2\pi ijq/a} \sum_{n=-\infty}^{\infty}
\delta(p-p_0-nb),
\label{classef}
\end{eqnarray}
with associated eigenvalues
\begin{equation}
\lambda^{(c)}_{j,p_0} = 2 \pi j p_0/\mu a
\label{classev}
\end{equation}
with $0 \le p_0<b$. Here the second identity results from an application
of Eq. (\ref{id1}).

A comparison of the classical [Eqs. (\ref{classef}) and (\ref{classev})]
and quantum [Eqs. (\ref{eq31}) and (\ref{eigene})] results shows that the
quantum and classical eigenfunctions are identical and the quantum
eigenvalues go to the proper classical limit as $h \rightarrow 0$.
This is not the case with the results of the traditional quantization
procedure, as discussed by Ford\cite{ford}.

\subsection{ Remarks on Extended Quantization}

The procedure above involves two modifications to the standard quantization
approach: (a) the introduction of modular variables and operators into
both the classical and quantum problem and (b) the use of the
von Neumann equation in quantum mechanics. The need for the latter stems from
the fact that the solutions which we find are, in fact, not pure states
and are not comprised of averages over pure states. That is, we find that
solutions to the von Neumann equation exist where Hamiltonian
eigenfunctions do not. The need for modular variables and operators is
also worth emphasizing.  Specifically, note that the solutions Eq.
(\ref{eigenf}) and Eq. (\ref{eigenfww}) are not eigenfunctions of their
respective quantum and classical Liouville operators if these operators
are written in terms of nonmodular $p,q$ variables. Indeed
eigenfunctions of the Liouville operators with Cartesian $p,q$ variables
and periodic boundary conditions
do not exist for mappings on the torus. This is easily seen as follows.
Consider the Liouville problem in $p,q$. Since the Hamiltonian in Eq. (1)
with $\epsilon=0$ is quadratic it gives rise to a
quantum Liouville operator which, in the Wigner-Weyl representation, is
identical to the classical Liouville operator . Thus, to obtain both the
quantum and classical Liouville eigenfunctions for $\epsilon=0$ we would
need only consider
the classical Liouville eigenequation $L_c\rho^{(c)}=\lambda^{c}\rho^{(c)}$
where
$L_c=-i(p/\mu)\partial /\partial q$. Taking matrix
elements in the Fourier basis would again yield Eq. (\ref{eige}) which as
we have seen has the solutions given by Eq. (\ref{eigenfww}). But
consideration of these solutions shows that they are eigensolutions of
$L_c = -i(p{\rm mod}~b/\mu)\partial /\partial q$, and not of $L_c =
-i(p/\mu)\partial /\partial q$.  Thus the procedure, starting with
Cartesian $p,q$ variables generates an inconsistency whose origin
is readily apparent. That is, although the Hamiltonian is of the
form given by Eq. (1), the Hilbert space chosen to solve the problem is
that spanned by the Fourier basis functions. But the Hamiltonian in
Eq. (1) and its associated Liouville operator cannot be decomposed on this
Fourier basis, i.e., they do not satisfy the translational invariances of
the Fourier basis. Hence, mathematical inconsistency and an incorrect form for
the quantum propagator result.

The advantages of the von Neumann approach over the conventional
Hamiltonian quantization are now evident.  Equations (\ref{eigenf}) and
(\ref{eigene}) provide a solution for any value of $ab$, are integrable
over any of the periodic unit cells and are well behaved.  Further, in the
classical limit ($h \rightarrow 0$), $\lambda_{j,p_0} \rightarrow
\lambda_{j,p_0}^{(c)}$  [compare Eqs. (\ref{eigene}) and (\ref{classev})]
so that the classical limit is properly reached by both eigenfunctions and
eigenvalues.

\section{Classical and Quantum Mapping Dynamics ($\epsilon \ne 0$)}

Here we focus on the
more general case of $\epsilon\neq 0$, obtain the propagator and show that
it readily yields the correct classical limit.

\subsection{Classical Mechanics}

Consider then the Hamiltonian [Eq. (1)] in reduced modular variables
($\eta=Tb/\mu a\in\bf{Z}$ and $\xi=-\epsilon Ta/b\in\bf{Z}$):
\begin{equation}
H(P,Q,t)=\frac{ab}{2T}\{\eta(P{\rm
mod}~1)^2-\xi(Q{\rm
mod}~1)^2\sum_{s=-\infty}^{\infty}\delta(s-t/T)\}.
\label{myh2}
\end{equation}
The classical dynamics of the map is defined by the classical propagator
\begin{equation}
\Lambda_c(T)={\cal T}\exp\{-i\int_{MT}^{(M+1)T}dtL_c(t)\}
\end{equation}
where $L_c(t)$ is the classical Liouville operator associated with Eq.
(\ref{myh2}) and where the integral is over a time interval beginning
just after the $M^{th}$ kick to just after the $(M+1)^{st}$ kick. Then
the dynamics is given by
\begin{equation}
\Lambda_c(T)\left( \begin{array}{c}
Q{\rm mod}~1 \\
P{\rm mod}~1
\end{array} \right)
=\phi^{-1}
\left( \begin{array}{c}
Q \\
P
\end{array} \right)~{\rm mod}~1
\end{equation}
where
\begin{equation}
\left( \begin{array}{c}
Q_{n+1} \\
P_{n+1}
\end{array} \right)
=\phi\left( \begin{array}{c}
Q_{n} \\
P_{n}
\end{array} \right){\rm mod}~1
=
\left( \begin{array}{cc}
1&\eta \\
\xi &1+\eta\xi
\end{array} \right)
\left( \begin{array}{c}
Q_n \\
P_n
\end{array} \right)~{\rm mod}~1.
\label{cmatrix}
\end{equation}
These linear mappings $\phi$ are such that
$I=\xi Q^2+\eta\xi QP-\eta P^2$
is a constant of the motion when $|2+\eta\xi| \leq 2$.
However, when $|2+\eta\xi| >2$, $I$ is a hyperbola
which is wrapped densely around the torus by the ``mod~1'' operation
and so effectively destroyed. We examine the quantum dynamics of the map
in both these regimes and obtain
results for all $\eta,\xi$, including the Arnold cat map\cite{arnold}
($\eta=\xi=1$), and systems like the quasiperiodic discretized particle in a
box\cite{dumont} ($\xi=0$).

To obtain the time evolution of the phase space density which at time zero
is $\rho^{(c)}(P,Q,0)$,
we Fourier expand in $f_{n,m}$. At time zero, and later time $MT$, we have:
\begin{equation}
\rho^{(c)}(P,Q,0)=\frac{1}{ab}\sum_{n,m=-\infty}^{\infty}
\rho^{(c)}_{n,m}(0) f_{n,m}
\label{eq48}
\end{equation}
\begin{equation}
\rho^{(c)}(P,Q,MT)=\frac{1}{ab}\sum_{n,m=-\infty}^{\infty}
\rho^{(c)}_{n,m}(MT) f_{n,m} =\Lambda_c(T)^M \rho^{(c)}(P,Q,0).
\label{eq54j}
\end{equation}

A straightforward analysis of the effect of $\Lambda_c(T)$ on
$\rho^{(c)}(P,Q,0)$ gives
\begin{equation}
\Lambda_c(T) \rho^{(c)}(P,Q,0)= \frac{1}{ab}\sum_{n,m} \rho^{(c)}_{\phi'(nm)}
f_{n,m}
\label{eq55j}
\end{equation}
where $\phi'(nm)$ is the transpose of the matrix in Eq. (\ref{cmatrix})
multiplying the column vector with elements $n,m$.
Clearly, evolution of the map, from this perspective, corresponds to the
interchange of coefficients in the Fourier basis $f_{n,m}$. We define
the classical propagator $G_c$ of the Fourier coefficients, for particular
reference
with the quantum result obtained later below, as
\begin{equation}
\rho^{(c)}_{n,m}(T) = \sum_{k,l=-\infty}^{\infty} G_c(n,m;k,l)
\rho^{(c)}_{k,l}(0).
\label{rhonmc}
\end{equation}
Comparison of Eq. (\ref{eq54j}) for $M=1$ and Eq. (\ref{eq55j})
reveals that
\begin{equation}
G_c(n,m;k,l) = \delta_{(k,l), \phi'(nm)}.
\label{gc}
\end{equation}
Generalization to the step from time $MT$ to  time $(M+1)T$ gives
\begin{equation}
\rho^{(c)}_{n,m}((M+1)T) = \sum_{k,l}G_c(n,m;k,l)\rho^{(c)}_{k,l}(MT)
\label{ugh})
\end{equation}
so that
\begin{equation}
\rho^{(c)}(P,Q,(M+1)T) =
\frac{1}{ab}\sum_{n,m}[\sum_{k,l}G_c(n,m;k,l) \rho^{(c)}_{k,l}(MT)] f_{n,m}.
\label{fullcprop}
\end{equation}
Equation (\ref{fullcprop}) provides a general expression for the time
evolution of the classical phase space density in terms of the classical
propagator $G_c$ matrix elements as coefficients in a Fourier expansion.

\subsection{Quantum Mechanics}

Consider now the quantum dynamics with the Hamiltonian, the analog of Eq.
(\ref{myh2}) with the appropriate introduction of mod operators,
\begin{equation}
\hat{H}(t)=\frac{ab}{2T}\{\eta(\widehat{P{\rm
mod}~1})^2-\xi(\widehat{Q{\rm mod}~1})^2\sum_{s=-\infty}^{\infty}\delta(
s-t/T)\}.
\label{qmyh2}
\end{equation}
The associated time evolution operator is
\begin{equation}
\hat{U}(T)={\cal T}e^{-i\int_{MT}^{(M+1)T}dt\hat{H}(t)/\hbar}
\end{equation}
where the integration is performed from just after the $M^{th}$ kick to just
after the $(M+1)^{st}$ kick. The resultant $\hat{U}(T)$ is $M$ independent.

The goal of our calculation below is to obtain the quantum analog of
$G_c$. To do so is complicated in execution but simple in its essence.
Specifically, we introduce the Fourier operator basis [Eq. (\ref{qfb})]
for the quantum distribution
$\hat{\rho}$ at time $MT$ as
\begin{equation}
\hat{\rho}(MT) = \frac{1}{ab} \sum_{n,m=-\infty}^{\infty}
\rho_{n,m}(MT)\hat{f}_{n,m}
\label{eq56}
\end{equation}
and endeavor to express the time evolution of the Fourier coefficients
$\rho_{n,m}$ as	the quantum analog of Eq. (\ref{ugh}), i.e. as
\begin{equation}
\rho_{n,m}((M+1)T) =
\sum_{k,l=-\infty}^{\infty}G(n,m;k,l)\rho_{k,l}(MT).
\label{fourprop}
\end{equation}
Doing so will allow an analysis of the quantum dynamics in the Fourier
basis as well as a direct comparison with the classical result. Indeed
this comparison with the classical limit benefits considerably from use of
the Wigner-Weyl representation. In this representation Eq. (\ref{eq56})
assumes the form
\begin{equation}
\rho(P,Q,MT) = \frac{1}{ab} \sum_{n,m=-\infty}^{\infty}
\rho_{n,m}(MT)f_{n,m}
\label{rhomt}
\end{equation}
so that, given Eq. (\ref{fourprop}),
\begin{equation}
\rho(P,Q,(M+1)T) =
\frac{1}{ab}\sum_{n,m}[\sum_{k,l}G(n,m;k,l) \rho_{k,l}(MT)] f_{n,m}.
\label{rhomtp}
\end{equation}
A comparison of Eq. (\ref{rhomtp}) with Eq. (\ref{fullcprop}) shows that
classical-quantum correspondence requires\cite{expect}
\begin{equation}
\lim_{\alpha \rightarrow 0} \rho_{k,l}(MT) = \rho^{(c)}_{k,l}(MT)
\label{req1}
\end{equation}
and
\begin{equation}
\lim_{\alpha \rightarrow 0}  G(n,m;k,l) = G_c(n,m;k,l).
\label{req2}
\end{equation}
Equation (\ref{req1}) holds if Eq. (\ref{req2}) holds and if
$\rho_{k,l}(0) = \rho^{(c)}_{k,l}(0)$. However, the latter equality
is assured, as seen by comparing
Eq. (\ref{rhomt}) with Eq. (\ref{eq48}), at time zero. Thus to prove the
correspondence limit for the fundamental dynamical entity
$\rho(P,Q,MT)$ requires that we show Eq. (\ref{req2}). This is
done in Section V after we obtain a general expression for the quantum
$G(n,m;k,l)$. Readers uninterested in the details of this derivation
should proceed to Eq. (\ref{gprop}).

\subsubsection{Quantum Propagator}

The time evolved density operator satisfies
\begin{equation}
\hat{\rho}((M+1)T)=\hat{U}(T)\hat{\rho}(MT)\hat{U}^{\dag}(T).
\label{ted}
\end{equation}
We transform to the Wigner-Weyl representation
and focus on the initial step from
time 0 to $T$; the same algebra pertains to the general step from
time $MT$ to $(M+1)T$.
Equation (\ref{ted}) in the Wigner-Weyl representation is
\begin{equation}
\rho(P,Q,T)=\frac{1}{ab}
\sum_{n,m=-\infty}^{\infty}~\rho_{n,m}~\int_{-\infty}^{\infty}dv
e^{i2\pi PV/\alpha}\langle
q-v/2|\hat{K}\hat{U}_0(T)\hat{f}_{n,m}\hat{U}_0^{\dag}(T)\hat{K}^{\dag}|q+v/2\
\rangle
\label{w1}
\end{equation}
where
\begin{eqnarray}
\hat{K}&=&\exp\{i\pi\xi(\widehat{Q{\rm mod}~1})^2/\alpha\}
\nonumber \\
\hat{U}_0(T)&=&\exp\{-i\pi\eta(\widehat{P{\rm
mod}~1})^2/\alpha\},
\end{eqnarray}
so that $\hat{U}(T) = \hat{K}\hat{U}_0$. [ Here again we employ a
convention where capital letters denote scaled variables i.e. $P=p/b$,
$Q=q/a$, $V=v/a$ and below $X=x/a$.]
The matrix element in Eq. (\ref{w1}) can be written as
\begin{eqnarray}
&&\langle
q-v/2|
\hat{K}\hat{U}_0(T)\hat{f}_{n,m}\hat{U}_0^{\dag}(T)\hat{K}^{\dag}|q+v/2\rangle=
\nonumber
\\
&&\int_{-\infty}^{\infty}dq'~e^{2\pi imQ'}\langle
q-v/2|\hat{K}\hat{U}_0(T)|q'-n\alpha a/2\rangle\langle q'+n\alpha
a/2|\hat{U}_0^{\dag}(T)\hat{K}^{\dag}|q+v/2\rangle,
\label{mel}
\end{eqnarray}
which involves matrix elements over $\hat{K}\hat{U}_0(T)$.
These matrix elements may be rewritten as (see Appendix B)
\begin{equation}
\langle x|\hat{K}\hat{U}_0(T)|x^{\prime}\rangle
=\exp\{i\pi\xi(X{\rm
mod}~1)^2/\alpha\}\sum_{k=-\infty}^{\infty}g_k(\eta)\delta(x-x^{\prime}-k\alpha
a),
\label{formu}
\end{equation}
where
\begin{equation}
g_k(z)=\int_0^{1}d\nu~e^{-i\pi z\nu^2/\alpha}e^{2\pi i
k\nu}.
\label{gee}
\end{equation}
Inserting Eq. (\ref{formu}) into Eq. (\ref{mel}) gives
\begin{eqnarray}
&&\langle
q-v/2|
\hat{K}\hat{U}_0(T)\hat{f}_{n,m}\hat{U}_0^{\dag}(T)\hat{K}^{\dag}|q+v/2\rangle=
\nonumber
\\
&&\sum_{k,l=-\infty}^{\infty}g_k(\eta)g_l^*(\eta)e^{i\pi\xi[(Q-V/2){\rm
mod}~1]^2/\alpha}e^{-i\pi\xi[(Q+V/2){\rm
mod}~1]^2/\alpha}\cdot \nonumber \\
&&e^{2\pi
im(Q-(k+l)\alpha/2)}\delta(v-(n+l-k)\alpha a).
\label{mel2}
\end{eqnarray}
Inserting Eq. (\ref{mel2}) into Eq. (\ref{w1}) gives, after some
algebra
\begin{equation}
\rho(P,Q,T)=\frac{1}{ab}\sum_{n,m=-\infty}^{\infty}
\rho_{n,m}(0)e^{2\pi
i(nP+mQ)}\kappa_{n,m}(P,Q)
\label{w2}
\end{equation}
with
\begin{eqnarray}
\kappa_{n,m}(P,Q)
&=&\sum_{k,l=-\infty}^{\infty}g_k(\eta)g_l^*(\eta)
e^{2\pi i(l-k)P}\cdot \nonumber \\
&&e^{-i\pi\alpha(k+l)m}e^{i\pi\xi[(Q-\alpha(n-k+l)/2){\rm
mod}~1]^2/\alpha}e^{-i\pi\xi[(Q+\alpha(n-k+l)/2){\rm mod}~1]^2/\alpha}.
\label{k0}
\end{eqnarray}
Equations (\ref{w2}) and (\ref{k0}) provide an expression for propagation
of the quantum map in phase space. However, extensive algebraic manipulation
is necessary to extract the propagator in the Fourier basis. We
begin by noting that we can write
\begin{equation}
\rho_{n,m}=\int_0^{b}dp_0\int_0^adq_0~\rho(P_0,Q_0,0)
e^{-2\pi i(nP_0+mQ_0)}
\end{equation}
so that Eq. (\ref{w2}) can be written in the form
\begin{equation}
\rho(P,Q,T)=\int_0^{b}dp_0\int_0^adq_0~
\rho(P_0,Q_0,0)\kappa(P,Q,T;P_0,Q_0,0)
\label{w3}
\end{equation}
where
\begin{equation}
\kappa(P,Q,T;P_0,Q_0,0)=
\frac{1}{ab}\sum_{n,m=-\infty}^{\infty}~e^{2\pi
i[n(P-P_0)+m(Q-Q_0)]}
\kappa_{n,m}(P,Q)
\label{k1}
\end{equation}
is the kernel of the Liouville propagator in the Wigner-Weyl
representation.	Since
\begin{equation}
\rho(P,Q,T)= \frac{1}{ab}\sum_{n,m} \rho_{nm}(T)
e^{2\pi i(nP+mQ)}
\label{old42}
\end{equation}
the Fourier coefficients of the propagated Wigner functions are
\begin{equation}
\rho_{n,m}(T)=\int_0^{b}dp_0\int_0^adq_0~\rho(P_0,Q_0,T)e^{-2\pi
i(nP_0+mQ_0)}.
\end{equation}
Given Eqs. (\ref{w3}) - (\ref{old42}), the goal is to rewrite the
propagation of $\rho_{n,m}(0)$ in terms of Eq. (\ref{fourprop}),
so as to extract $G(n,m;k,l)$, i.e. in the first iteration we want
\begin{equation}
\rho_{n,m}(T)=\sum_{k,l=-\infty}^{\infty}G(n,m;k,l)\rho_{k,l}(0).
\label{eq72}
\end{equation}
We begin by noting that Eq. (\ref{k1}) can be manipulated, in accord with
Appendix C, into the form
\begin{eqnarray}
&&\kappa(P,Q,T;P_0,Q_0,0)=\frac{1}{ab}\sum_{n,m=-\infty}^{\infty}~e^{2\pi
i[nP+mQ]}\cdot \nonumber \\
&&\sum_{k,l=-\infty}^{\infty}g_k(\eta)g_l^*(\eta)
e^{-2\pi i[(n+k-l)P_0+mQ_0]}
e^{-i\pi\alpha(k+l)m}\cdot \nonumber \\
&& e^{i\pi\xi[(Q_0-\alpha(n-k-l)/2){\rm
mod}~1]^2/\alpha}e^{-i\pi\xi[(Q_0+\alpha(n+k+l)/2){\rm
mod}~1]^2/\alpha}
\label{k2}
\end{eqnarray}
and so Eq. (\ref{w3}) can be written as
\begin{eqnarray}
&&\rho(P,Q,T)=\frac{1}{ab}\sum_{n,m=-\infty}^{\infty}~e^{2\pi
i[nP+mQ]}
\int_0^{b}dp_0\int_0^adq_0~\rho(P_0,Q_0,0) \nonumber \\
&&\sum_{k,l=-\infty}^{\infty}g_k(\eta)g_l^*(\eta)
e^{-2\pi i[(n+k-l)P_0+mQ_0]}
e^{-i\pi\alpha(k+l)m}\cdot \nonumber \\
&& e^{i\pi\xi[(Q_0-\alpha(n-k-l)/2){\rm
mod}~1]^2/\alpha}e^{-i\pi\xi[(Q_0+\alpha(n+k+l)/2){\rm
mod}~1]^2/\alpha}.
\label{w4}
\end{eqnarray}
Comparing Eq. (\ref{w4}) to Eq. (\ref{old42}) we identify
\begin{eqnarray}
&&\rho_{n,m}(T)=\int_0^{b}dp_0\int_0^adq_0~\rho(P_0,Q_0,0)\cdot \nonumber \\
&&\sum_{k,l=-\infty}^{\infty}g_k(\eta)g_l^*(\eta)
e^{-2\pi i[(n+k-l)P_0+mQ_0]}
e^{-i\pi\alpha(k+l)m}\cdot \nonumber \\
&&e^{i\pi\xi[(Q_0-\alpha(n-k-l)/2){\rm
mod}~1]^2/\alpha}e^{-i\pi\xi[(Q_0+\alpha(n+k+l)/2){\rm
mod}~1]^2/\alpha}.
\end{eqnarray}
Using the identity (see Appendix B for a proof)
\begin{equation}
e^{-i\pi z(x{\rm
mod}~1)^2/\alpha}=\sum_{j=-\infty}^{\infty}~g_j(z)~e^{-2\pi ixj},
\label{id2}
\end{equation}
one obtains
\begin{eqnarray}
\rho_{n,m}(T)&=&\sum_{i,j,k,l=-\infty}^{\infty}g_i(\xi)g_j^*(\xi)g_k(\eta)g_l^*(\eta)\cdot\nonumber
\\
&&\rho_{n+k-l,m+i-j}(0)e^{-i\pi\alpha(k+l)(m+i-j)}e^{-i\pi\alpha(i+j)n}.
\label{ans1}
\end{eqnarray}
This can immediately be generalized to the iterative form giving
$\rho_{n,m}((M+1)T)$ in terms of $\rho_{n,m}$ at time $MT$.
However, although this equation correctly describes the manner in which the
Fourier coefficients of a Wigner distribution are mapped under the
quantum Liouville propagator it is too complicated to be of use.
Let $k^{\prime}=n+k-l$ and $i^{\prime}=m+i-j$. Under this change in
the integers of summation Eq. (\ref{ans1}) becomes
\begin{eqnarray}
\rho_{n,m}(T)&=&\sum_{i^{\prime},j,k^{\prime},l=-\infty}^{\infty}g_{i^{\prime}+j-m}(\xi)g_j^*(\xi)g_{k^{\prime}+l-n}(\eta)g_l^*(\eta)\cdot\nonumber
\\
&&\rho_{k^{\prime},i^{\prime}}(0)e^{-
i\pi\alpha(k^{\prime}+2l-n)i^{\prime}}e^{-
i\pi\alpha(i^{\prime}+2j-m)n}.
\label{ans2}
\end{eqnarray}
But the sums over $j$ and $l$ may be done explicitly. To begin with
\begin{eqnarray}
&&\sum_{j=-\infty}^{\infty}g_j^*(\xi)g_{i^{\prime}+j-m}(\xi)e^{-2\pi
i(\alpha
n)j}=\sum_{j=-\infty}^{\infty}\int_0^1d\nu\int_0^1d\nu^{\prime}\cdot
\nonumber \\
&&e^{i\pi\xi\nu^2/\alpha}e^{-
2\pi i j\nu}e^{-i\pi\xi\nu^{\prime 2}/\alpha}e^{2\pi
i(i^{\prime}+j-m)\nu^{\prime}}e^{-2\pi i(\alpha n)j}
\end{eqnarray}
and use of identity (\ref{id1}) yields
\begin{eqnarray}
&&\sum_{j=-\infty}^{\infty}g_j^*(\xi)g_{i^{\prime}+j-m}(\xi)e^{-2\pi
i(\alpha
n)j}=\int_0^1d\nu\int_0^1d\nu^{\prime}
e^{i\pi\xi\nu^2/\alpha}\cdot\nonumber \\
&&e^{-i\pi\xi\nu^{\prime
2}/\alpha}e^{2\pi
i(i^{\prime}-m)\nu^{\prime}}\delta(\nu^{\prime}-(\nu+\alpha n){\rm
mod}~1) \\
&=&\int_0^1d\nu e^{-i\pi\xi[(\nu+\alpha
n){\rm mod}~1]^2/\alpha}
e^{i\pi\xi\nu^2/\alpha}e^{2\pi
i(i^{\prime}-m)(\nu+\alpha n)}.
\end{eqnarray}
Similar calculations give
\begin{equation}
\sum_{l=-\infty}^{\infty}g_l^*(\xi)g_{k^{\prime}+l-n}(\xi)e^{-2\pi
i(\alpha
n)j}=\int_0^1d\nu^{\prime} e^{-i\pi\xi[(\nu^{\prime}+\alpha
i^{\prime}){\rm mod}~1]^2/\alpha}
e^{i\pi\xi\nu^{\prime 2}/\alpha}e^{2\pi
i(k^{\prime}-n)(\nu^{\prime}+\alpha i^{\prime})}.
\end{equation}
Insertion of these results into Eq. (\ref{ans2}), setting
$k^{\prime}=k$ and $i^{\prime}=l$, and simplifying, gives the desired
result, i.e. Eq. (\ref{eq72}) with
\begin{eqnarray}
&&G(n,m;k,l)=e^{
i\pi\alpha(kl-nm)}\int_0^1d\nu\int_0^1d\nu^{\prime}e^{-i\pi\xi[(\nu+\alpha
n){\rm mod}~1]^2/\alpha}e^{i\pi\xi\nu^2/\alpha}\cdot\nonumber \\
&&e^{2\pi i(l-m)\nu}
e^{-i\pi\eta[(\nu^{\prime}+\alpha
l){\rm mod}~1]^2/\alpha}e^{i\pi\eta\nu^{\prime 2}/\alpha}e^{2\pi
i(k-n)\nu^{\prime}}
\label{gprop}
\end{eqnarray}
as the propagator of the Fourier coefficients\cite{mkick}.
This can be immediately generalized to give Eq. (\ref{fourprop})
with the same propagator $G$.
The crucial difference between
the classical $G_c$ [Eqs.(\ref{rhonmc}),(\ref{gc})] and the quantum propagator
$G$ is clear insofar as the quantum propagator mixes
contributions from all $k,l$ components to produce each $n,m$ whereas
$G_c$ does not.

Note that propagator (\ref{gprop}) moves distributions forward in time. To
obtain its inverse, i.e. the propagator which moves distributions
backward in time, simply take the complex conjugate of Eq.
(\ref{gprop}) and interchange $n$ with $k$, and $m$ with $l$ (see
Appendix D for a proof).

As a simple check on
the validity of the quantization procedure we redo the ``free
particle" case i.e. $\xi=0$. In this case $G$ is given by
\begin{equation}
G(n,m;k,l)=\delta_{l,m}e^{i\pi\alpha
m(k-n)}\int_0^1d\nu e^{-i\pi\eta[(\nu+\alpha m){\rm
mod}~1]^2/\alpha}e^{i\pi\nu^2/\alpha}e^{2\pi i(k-n)\nu}
\end{equation}
and the eigenequation is
\begin{equation}
\sum_{k,l=-\infty}^{\infty}G(n,m;k,l)\rho_{k,l}=e^{-i\lambda T}\rho_{n,m}.
\end{equation}
Consider a solution of the form
$\rho_{k,l}=\Omega_le^{-ik\theta_l}$. Substitution and use of
Eq. (\ref{id1}) yields
\begin{equation}
e^{-i\pi\alpha mn}\int_0^1d\nu e^{-i\pi\eta[(\nu+\alpha m){\rm
mod}~1]^2/\alpha}e^{i\pi\nu^2/\alpha}e^{-2\pi
in\nu}\delta(\nu-(\frac{\theta_m}{2\pi}-\alpha m/2){\rm
mod}~1)=e^{-i\lambda T}e^{-in\theta_m}
\end{equation}
and this implies that
\begin{equation}
\lambda =\frac{\pi\eta}{\alpha T}\{[(\frac{\theta_m}{2\pi}+\alpha m/2){\rm
mod}~1]^2-[(\frac{\theta_m}{2\pi}-\alpha m/2){\rm mod}~1]^2\}.
\end{equation}
Choosing $\frac{\theta_m}{2\pi}=\frac{p_0}{b}$ where $p_0\in[0,b)$
gives the same solutions [Eqs. (\ref{eigenf}) and (\ref{eigene})]
which we obtained previously.

\section{Correspondence}

Given Eq. (\ref{gprop}) it is straightforward to examine the dynamics in the
classical limit ($\alpha \rightarrow 0)$. We do so at fixed
finite time so that, with respect to the issue of ergodic properties, we
are taking the correct order of limits, $h \rightarrow 0$ prior to any
long time limit\cite{berry2}.

For each set of values $\{\nu,\nu^{\prime},n,l\}$ there exists a
sufficiently small $\alpha$ such that
$(\nu+\alpha n){\rm mod}~1=\nu+\alpha n$ and
$(\nu^{\prime}+\alpha l){\rm mod}~1=\nu^{\prime}+\alpha l$.
Thus in the $\alpha\rightarrow 0$ limit
\begin{eqnarray}
G(n,m;k,l)\rightarrow &&\int_0^1d\nu\int_0^1d\nu^{\prime}e^{2\pi
i(l-m-\xi n)\nu}e^{2\pi i(k-n-\eta l)\nu^{\prime}} \nonumber \\
&=&\delta_{l,m+\xi n}\delta_{k,n+\eta l}
=\delta_{l,m+\xi n}\delta_{k,n+\eta m +\eta \xi n} \nonumber \\
&=&\delta_{(l,k),\phi^{\prime}\cdot(n,m)}
\end{eqnarray}
which is the correct classical limit, Eq. (\ref{gc}). Note that this result
holds for all maps of the form given by Eq. (\ref{cmatrix}), including the
classical chaotic cat map.

Having demonstrated that these quantum maps give the correct classical
limit it is necessary to obtain its dynamical characteristics in the
case of $h \ne 0$. It is straightforward to demonstrate that,
independent of the value of $\eta,\xi$, the
quantum propagator has at least two eigenfunctions with unit
eigenvalue and hence
that the quantum dynamics is nonergodic. Two such eigenfunctions are the
uniform distribution and the Schrodinger propagator. To see this consider
the uniform distribution
$1_{k,l}(0)=\delta_{k,0}\delta_{l,0}$ under propagation.
Using Eq.
(\ref{gprop}) we have
\begin{eqnarray}
1_{n,m}(T)&=&G(n,m;0,0) \nonumber \\
&=&e^{-2\pi i mn\alpha}\int_0^1d\nu\int_0^1d\nu^{\prime}
e^{-i\pi\xi[(\nu+\alpha
n){\rm mod}~1]^2/\alpha}e^{i\pi\xi\nu^2/\alpha}e^{-2\pi
im\nu}e^{-2\pi in\nu^{\prime}} \nonumber \\
&=&\delta_{n,0}\int_0^1d\nu~e^{-2\pi
im\nu} \nonumber \\
&=&\delta_{n,0}\delta_{m,0}.
\end{eqnarray}
Thus the uniform distribution is an eigenfunction with unit eigenvalue.
[This result also implies that the dynamics is measure (i.e. area)
preserving.]

Second, note that the
Schrodinger propagator is an eigenfunction of the Liouville propagator since
$\hat{U}^{\dag}(\hat{U})\hat{U}=\hat{U}^{-1}(\hat{U})\hat{U}=\hat{U}$.
It remains to demonstrate that the Wigner function associated with
$\hat{U}$ is $L^2$ on a unit cell. To see this
note the form of $\hat{U}$ in the Wigner-Weyl representation:
\begin{equation}
U^{(w)}(P,Q)=\sum_{n,m=-\infty}^{\infty}g_{-n}(\eta)g_m^*(\xi)e^{-i\pi\alpha
nm}e^{2\pi i(nP+mQ)}.
\end{equation}
Then
\begin{eqnarray}
\int_0^1dP\int_0^1dQ|U^{(w)}(P,Q)|^2
&=&\int_0^1dP\int_0^1dQ
\sum_{n,m,k,l=-\infty}^{\infty}g_{-k}^*(\eta)g_l(\xi)g_{-n}(\eta)g_m^*(\xi)\
\cdot
\nonumber \\
&&e^{2\pi
i[(n-k)P+(m-l)Q]}e^{-i\pi(nm-kl)\alpha} \\
&=&\sum_{k=-\infty}^{\infty}|g_k(\eta)|^2\cdot\sum_{l=-\infty}^{\infty}|g_l(\xi) |^2 \\
&=&1^2=1
\end{eqnarray}
since
\begin{equation}
\sum_{k=-\infty}^{\infty}g_{j+k}(z)g_k^*(z)=\delta_{j,0}.
\label{id3}
\end{equation}
The latter identity is proven in Appendix B.

Thus $U^{(w)}$ is an $L^2$ eigenfunction of the mapping with
eigenvalue 1, as is the uniform distribution and so\cite{arnold} the mappings
are not ergodic. Note, furthermore, the $\alpha\rightarrow 0$ limit of
$U^{(w)}$ displays an essential singularity. Hence
this eigenfunction does not exist classically, a result which is essential
to achieve the proper ergodic classical limit for classically ergodic
cases like the cat map.

The extended quantization
procedure therefore provides a straightforward treatment of the cat map, a
system which has been the subject of considerable controversy.

\section{Why No Hamiltonian Eigenfunctions?}

In addition to the explicit results on maps, this work has also introduced
a quantization procedure leading to results which differ significantly from the
traditional Schrodinger equation approach. This difference arises
primarily from the fact that periodic boundary conditions applied
to the density matrix [$T^{-1} \hat{\rho} T= \hat{\rho}$ for $T=T_q(a)$
and $T=T_p(b)$] allow for solutions when similar boundary
conditions, applied to the wavefunction, preclude the possibility of a
solution. Formally this is so because periodic boundary conditions on the
wavefunctions (corresponding to the condition $T \hat{\rho}= \hat{\rho},
\hat{\rho} T = \hat{\rho}$, on the density matrix) is a
far more restrictive requirement than the condition $T^{-1} \hat{\rho} T=
\hat{\rho}$. Nonetheless it is of interest to examine the character of the
mapping
dynamics and to expose the physics underlying the lack of wavefunctions.

Consider the Arnold cat map\cite{arnold} as a specific example.
Written explicitly, the cat map ($\eta = \xi =1$ in Eq. (\ref{cmatrix}) reads
\begin{eqnarray}
Q_{n+1}{\rm mod}~1-Q_{n}{\rm mod}~1&=&P_{n}{\rm
mod}~1+V(Q_{n}{\rm mod}~1,P_{n}{\rm mod}~1)
\label{map1} \\
P_{n+1}{\rm mod}~1-P_n{\rm mod}~1&=&P_n{\rm
mod}~1+Q_n{\rm mod}~1+F(Q_n{\rm mod}~1,P_n{\rm
mod}~1)
\label{map2}
\end{eqnarray}
where
\begin{eqnarray}
V(Q{\rm mod}~1,P{\rm mod}~1)&=&-1\chi_{A_3\cup
A_4}(Q{\rm mod}~1,P{\rm mod}~1), \\
F(Q{\rm mod}~1,P{\rm mod}~1)&=&-1\chi_{A_2\cup
A_3}(Q{\rm mod}~1,P{\rm mod}~1)
-2\chi_{A_4}(Q{\rm mod}~1,P{\rm mod}~1).
\end{eqnarray}
Here $\chi_B$ is the characteristic function on the set $B$ and the
phase space regions $A_i$ are defined by:
\begin{eqnarray}
A_2&=&\{\frac{1}{2}\leq Q{\rm mod}~1<
1\}\times\{1-2Q{\rm mod}~1\leq
P{\rm mod}~1<1-Q{\rm mod}~1\} \\
A_3&=&\{0\leq Q{\rm mod}~1< \frac{1}{2}\}\times\{1-Q{\rm mod}~1\
\leq
P{\rm mod}~1<2-2Q{\rm mod}~1\} \\
A_4&=&\{\frac{1}{2}\leq Q{\rm mod}~1<
1\}\times\{2-2Q{\rm mod}~1\leq
P{\rm mod}~1<1\}.
\end{eqnarray}
Equations (\ref{map1}) and (\ref{map2}) show
that the changes in $q$ and $p$ under the mapping depend upon an
external boost $V$ and external force $F$ which are functions of position and
momentum. Thus the cat map equations
are a discretization of a Langevin type process
\begin{eqnarray}
\frac{dq}{dt}&=&p/\mu +V(q,p) \nonumber\\
\frac{dp}{dt}&=&-\zeta p-\frac{\partial \Phi}{\partial q}+F(p,q)
\label{Lang}
\end{eqnarray}
with a negative coefficient of friction $\zeta=-1$ and potential
$\Phi=-q^2/2$. Since
this set of equations describes the Hamiltonian dynamics
of a particle moving in the vicinity of a single hyperbolic
point if $V=\zeta=F=0$, Eqs. (44) and (45) constitute a set of perturbed
dynamics around an unstable point.
Thus the classical cat map can be regarded as a system subject to
external boosts and forces, rather than a stroboscopic viewing of
a true Hamiltonian system. Since the system is one which is subjected to
external boosts and forces, it does not admit a pure state description.

\section{Summary}

In summary, we have introduced a von Neumann based quantization procedure
in concert with Bloch type boundary conditions in order to achieve a
physically complete quantization of mappings on a torus\cite{commrel}$^{b}$.
The more
traditional approach\cite{ford,berry} of quantizing the mappings via the
Schrodinger equation with associated periodic boundary conditions yields
solutions only for tori satisfying $ab=hN$, and excessively restricts the
mapping dynamics to the dynamics on a set of rational points in $(q,p)$
space, associated with a discrete Liouville spectrum. Alternative
approaches such as applying Bloch type boundary conditions on the
wavefunction or periodic boundary conditions on the density matrix
encounter similar difficulties.  From the perspective of the density
matrix periodic boundary conditions imply  $T \hat{\rho}= \hat{\rho},
\hat{\rho} T = \hat{\rho}$ for $T=T_q(a)$ and $T=T_p(b)$.  Adopting Bloch
boundary conditions on the density matrix ($T^{-1} \hat{\rho}
T=\hat{\rho}$), the only remaining possibility, allows for a broader class
of solutions obtainable even when wavefunctions do not exist. The fact
that pure states do not in general exist is a consequence of the
fact that boundary conditions here correspond to the
application of external forces to the system.

The resultant quantum system has a continuous Liouville spectrum, and
the von Neumann quantized cat map dynamics is not ergodic until the proper
classical limit is reached.
This result provides the first demonstration of a straightforward limiting
procedure which completely recovers the full chaotic classical dynamics in
the $h \rightarrow 0$ limit. Further studies are in
progress to ascertain the character of the quantum dynamics and its
dependence on $h$ as the system approaches the classical limit.
Further, this study suggests the possibility of quantization via a
generalized von Neumann (or continuity) equation for systems described as
a Langevin type process i.e. Eqs. (\ref{Lang}), a fertile general area for
further study.

\vspace{0.2in}

\noindent
{\bf Acknowledgment:} Support from the Centre of Excellence in Molecular and
Interfacial Dynamics, and NSERC, is gratefully acknowledged.
J.W. acknowledges postgraduate fellowship support from NSERC and the OGS
program.

\pagebreak

\section*{Appendix A}

Here we discuss briefly the algebra generated by the operators
$\hat{f}_{n,m}$. It is trivial to show that these operators satisfy the
following conditions; (1) closure
$\hat{f}_{n,m}\hat{f}_{k,l}=e^{i\pi(nl-mk)\alpha}\hat{f}_{n+k,m+l}$, (2)
associativity $(\hat{f}_{n,m}\hat{f}_{k,l})\hat{f}_{i,j}=
\hat{f}_{n,m}(\hat{f}_{k,l}\hat{f}_{i,j})$, (3) the existence of an
identity $\hat{f}_{0,0}=\hat{1}$, (4) the existence of inverses
$\hat{f}_{n,m}\hat{f}_{-n,-m}=\hat{1}=\hat{f}_{-n,-m}\hat{f}_{n,m}$, and
(5) $\hat{f}_{n,m}^{\dag}=\hat{f}_{-n,-m}=(\hat{f}_{n,m})^{-1}$ which
implies that they are unitary operators.  Furthermore, 6) every element of
this algebra can be generated from the set
$\{\hat{f}_{1,0},~\hat{f}_{-1,0},~\hat{f}_{0,1},~\hat{f}_{0,-1}\}$.

\section*{Appendix B}

Here we will prove some of the simple formulae used in this paper.
Underlying all of these formulae is the identity in Eq. (\ref{id1}). We have
defined integrals
$g_k(z)=\int_0^{\infty}d\nu e^{-i\pi z\nu^2/\alpha}e^{2\pi ik\nu}$
and here we will prove two simple identities and one formula specific
to the quantum map problem involving these integrals.

The first identity is that of Eq. (\ref{id2}).
Using identity (\ref{id1})
\begin{eqnarray}
e^{-i\pi z(x{\rm
mod}~1)^2/\alpha}&=&\int_0^1d\nu\delta(\nu-x) e^{-i\pi z\nu^2/\alpha}
\nonumber \\
&=&\int_0^1d\nu\sum_{j=-\infty}^{\infty}e^{2\pi
i(\nu-x)j}e^{-i\pi z\nu^2/\alpha} \nonumber \\
&=&\sum_{j=-\infty}^{\infty}g_j(z)e^{-2\pi ijx}.
\end{eqnarray}

The second identity is that of Eq. (\ref{id3}).
Again using identity (\ref{id1})
\begin{eqnarray}
\sum_{j=-\infty}^{\infty}g_{j+k}(z)g_j^*(z)&=&
\int_0^1d\nu\int_0^1d\nu^{\prime}e^{-i\pi z\nu^2/\alpha}e^{i\pi z\nu^{\prime
2}/\alpha}e^{2\pi ik\nu}\sum_{j=-\infty}^{\infty}e^{2\pi
ij(\nu-\nu^{\prime})} \nonumber \\
&=&\int_0^1d\nu\int_0^1d\nu^{\prime}e^{-i\pi z\nu^2/\alpha}e^{i\pi z\nu^{\prime
2}/\alpha}e^{2\pi ik\nu}\delta(\nu-\nu^{\prime}) \nonumber \\
&=&\int_0^1d\nu e^{2\pi ik\nu} =\delta_{k,0}.
\end{eqnarray}

The formula specific to the quantum map problem is Eq. (\ref{formu}).
The proof is as follows;
\begin{eqnarray}
\langle
x|\hat{K}\hat{U}_0(T)|x^{\prime}\rangle&=&e^{i\pi\xi(X{\rm
mod}~1)^2/\alpha}\int_{-\infty}^{\infty}dp\langle x|p\rangle\langle
p|x^{\prime} \rangle e^{-i\pi\eta(P{\rm
mod}~1)^2/\alpha}\nonumber \\
&=&e^{i\pi\xi(X{\rm
mod}~1)^2/\alpha}\int_{-\infty}^{\infty}\frac{dp}{2\pi\hbar}e^{ip(x-x^{\prime})/\hbar}e^{-i\pi\eta(P{\rm
mod}~1)^2/\alpha} \nonumber \\
&=&e^{i\pi\xi(X{\rm
mod}~1)^2/\alpha}
\sum_{k=-\infty}^{\infty}g_k(\eta)\int_{-\infty}^{\infty}\frac{dp}{2\pi\hbar}
e^{ip(x-x^{\prime}-\alpha
ak)/\hbar} \nonumber \\
&=&e^{i\pi\xi(X{\rm
mod}~1)^2/\alpha}\sum_{k=-\infty}^{\infty}g_k(\eta)\delta(x-x^{\prime}-\alpha
ak),
\end{eqnarray}
where we have used identity (\ref{id2}) and the closure relation for
momentum.

\section*{Appendix C}

Here we will show that Eq. (\ref{k1}) can be put in the form of
Eq. (\ref{k2}). To begin, we substitute expression (\ref{k0}) for
$\kappa_{n,m}(P,Q)$ into Eq. (\ref{k1}). This
gives
\begin{eqnarray}
&&\kappa(P,Q,T;P_0,Q_0,0)=
\sum_{n,m=-\infty}^{\infty}~e^{2\pi
i[n(P-P_0)+m(Q-Q_0)]} \nonumber \\
&&\sum_{k,l=-\infty}^{\infty}g_k(\eta)g_l^*(\eta)
e^{2\pi i(l-k)P}\cdot \nonumber \\
&&e^{-
i\pi\alpha(k+l)m}e^{i\pi\xi[(Q-\alpha(n-k+l)/2){\rm
mod}~1]^2/\alpha}e^{-i\pi\xi[(Q+\alpha(n-k+l)/2){\rm
mod}~1]^2/\alpha}.
\end{eqnarray}
Using identity (\ref{id1}) for the sum over $m$ then gives
\begin{eqnarray}
&&\kappa(P,Q,T;P_0,Q_0,0)=
\sum_{n,m,k,l=-\infty}^{\infty}~e^{2\pi
in(P-P_0)}\delta(Q-Q_0-(k+l)\alpha/2
-m)\cdot\nonumber \\
&&g_k(\eta)g_l^*(\eta)
e^{2\pi i(l-k)P}e^{i\pi\xi[(Q-\alpha(n-k+l)/2){\rm
mod}~1]^2/\alpha}e^{-i\pi\xi[(Q+\alpha(n-k+l)/2){\rm
mod}~1]^2/\alpha}\\
&=&\sum_{n,m,k,l=-\infty}^{\infty}~e^{2\pi
in(P-P_0)}\delta(Q-Q_0-(k+l)\alpha/2
-m)\cdot\nonumber \\
&&g_k(\eta)g_l^*(\eta)
e^{2\pi i(l-k)P}e^{i\pi\xi[(Q_0-\alpha(n-2k)/2){\rm
mod}~1]^2/\alpha}e^{-i\pi\xi[(Q_0+\alpha(n+2l)/2){\rm
mod}~1]^2/\alpha}.
\end{eqnarray}
Now using identity (\ref{id1}) in reverse we obtain
\begin{eqnarray}
&&\kappa(P,Q,T;P_0,Q_0,0)=
\sum_{n,m,k,l=-\infty}^{\infty}~e^{2\pi
i[n(P-P_0)+m(Q-Q_0)]}
g_k(\eta)g_l^*(\eta)
e^{2\pi i(l-k)P}\cdot \nonumber \\
&&e^{-
i\pi\alpha(k+l)m}e^{i\pi\xi[(Q_0-\alpha(n-2k)/2){\rm
mod}~1]^2/\alpha}e^{-i\pi\xi[(Q_0+\alpha(n+2l)/2){\rm
mod}~1]^2/\alpha}.
\end{eqnarray}
Making the change in the integer of summation $n^{\prime}=n+l-k$ we
then obtain
\begin{eqnarray}
&&\kappa(P,Q,T;P_0,Q_0,0)=
\sum_{n^{\prime},m,k,l=-\infty}^{\infty}~e^{2\pi
i(n^{\prime}P+mQ)}
g_k(\eta)g_l^*(\eta)
e^{-2\pi i[(n^{\prime}+k-l)P_0+mQ_0]}\cdot \nonumber \\
&&e^{-
i\pi\alpha(k+l)m}e^{i\pi\xi[(Q_0-\alpha(n^{\prime}-k-l)/2){\rm
mod}~1]^2/\alpha}e^{-i\pi\xi[(Q_0+\alpha(n^{\prime}+k+l)/2){\rm
mod}~1]^2/\alpha}.
\end{eqnarray}
Now setting $n^{\prime}=n$ and changing the order of the factors we
obtain Eq. (\ref{k2}).

\section*{Appendix D}

Here we will show that taking the complex conjugate of propagator
(\ref{gprop}) and interchanging $n$ and $k$, and $m$ and $l$ gives the
inverse of propagator (\ref{gprop}). This inverse propagator
propagates distributions backward in time.
The inverse propagator according to this prescription is
\begin{eqnarray}
&&G^{-1}(n,m;k,l)=e^{
i\pi\alpha(kl-nm)}\int_0^1d\nu\int_0^1d\nu^{\prime}e^{i\pi\xi[(\nu+\alpha
k){\rm mod}~1]^2/\alpha}e^{-i\pi\xi\nu^2/\alpha}\cdot\nonumber \\
&&e^{2\pi i(l-m)\nu}
e^{i\pi\eta[(\nu^{\prime}+\alpha
m){\rm mod}~1]^2/\alpha}e^{-i\pi\eta\nu^{\prime 2}/\alpha}e^{2\pi
i(k-n)\nu^{\prime}}.
\end{eqnarray}
To verify that $G^{-1}$ is indeed the inverse of $G$ we need only show
that
\begin{equation}
\sum_{r,s=-\infty}^{\infty}G(n,m;r,s)G^{-1}(r,s;k,l)=\delta_{n,k}\delta_{m,l}=\sum_{r,s=-\infty}^{\infty}G^{-1}(n,m;r,s)G(r,s;k,l).
\label{identities}
\end{equation}
We begin with the first equality;
\begin{eqnarray}
&&\sum_{r,s=-\infty}^{\infty}G(n,m;r,s)G^{-1}(r,s;k,l)=e^{
i\pi(kl-nm)\alpha}\sum_{r,s=-\infty}^{\infty}\int_0^1d\nu\int_0^1d\nu^{\prime}\int_0^1d\nu^{\prime\prime}\int_0^1d\nu^{\prime\prime\prime}\cdot\nonumber
\\
&&e^{-i\pi\xi[(\nu+\alpha
n){\rm mod}~1]^2/\alpha}e^{i\pi\xi\nu^2/\alpha}
e^{2\pi i(s-m)\nu}
e^{-i\pi\eta[(\nu^{\prime}+\alpha
s){\rm mod}~1]^2/\alpha}e^{i\pi\eta\nu^{\prime 2}/\alpha}e^{2\pi
i(r-n)\nu^{\prime}}\cdot \nonumber \\
&&e^{i\pi\xi[(\nu^{\prime\prime}+\alpha
k){\rm mod}~1]^2/\alpha}e^{-i\pi\xi\nu^{\prime\prime 2}/\alpha}
e^{2\pi i(l-s)\nu^{\prime\prime}}
e^{i\pi\eta[(\nu^{\prime\prime\prime}+\alpha
s){\rm mod}~1]^2/\alpha}e^{-i\pi\eta\nu^{\prime\prime\prime 2}/\alpha}e^{2\pi
i(k-r)\nu^{\prime\prime\prime}}.
\end{eqnarray}
Using identity (\ref{id1}) to do the sum over $r$ gives
the delta function $\delta(\nu^{\prime\prime\prime}-\nu^{\prime})$ and
then doing the integral over $\nu^{\prime\prime\prime}$ we obtain
\begin{eqnarray}
&&\sum_{r,s=-\infty}^{\infty}G(n,m;r,s)G^{-1}(r,s;k,l)=e^{
i\pi(kl-nm)\alpha}\sum_{r,s=-\infty}^{\infty}\int_0^1d\nu\int_0^1d\nu^{\prime}\int_0^1d\nu^{\prime\prime}\cdot\nonumber
\\
&&e^{-i\pi\xi[(\nu+\alpha
n){\rm mod}~1]^2/\alpha}e^{i\pi\xi\nu^2/\alpha}
e^{2\pi i(s-m)\nu}
e^{-i\pi\eta[(\nu^{\prime}+\alpha
s){\rm mod}~1]^2/\alpha}e^{i\pi\eta\nu^{\prime 2}/\alpha}e^{2\pi
i(r-n)\nu^{\prime}}\cdot \nonumber \\
&&e^{i\pi\xi[(\nu^{\prime\prime}+\alpha
k){\rm mod}~1]^2/\alpha}e^{-i\pi\xi\nu^{\prime\prime 2}/\alpha}
e^{2\pi i(l-s)\nu^{\prime\prime}}
e^{i\pi\eta[(\nu^{\prime}+\alpha
s){\rm mod}~1]^2/\alpha}e^{-i\pi\eta\nu^{\prime 2}/\alpha}e^{2\pi
i(k-r)\nu^{\prime}}\\
&&=e^{
i\pi(kl-nm)\alpha}\sum_{r,s=-\infty}^{\infty}\int_0^1d\nu\int_0^1d\nu^{\prime}\int_0^1d\nu^{\prime\prime}\cdot\nonumber
\\
&&e^{-i\pi\xi[(\nu+\alpha
n){\rm mod}~1]^2/\alpha}e^{i\pi\xi\nu^2/\alpha}
e^{2\pi i(s-m)\nu}e^{2\pi
i(k-n)\nu^{\prime}}
e^{i\pi\xi[(\nu^{\prime\prime}+\alpha
k){\rm mod}~1]^2/\alpha}e^{-i\pi\xi\nu^{\prime\prime 2}/\alpha}
e^{2\pi i(l-s)\nu^{\prime\prime}}.
\end{eqnarray}
Now performing the sum over $s$, again through the use of identity
(\ref{id1}), gives the delta function $\delta(\nu^{\prime\prime}-\nu)$
and doing the integral over $\nu^{\prime\prime}$ we obtain
\begin{eqnarray}
&&\sum_{r,s=-\infty}^{\infty}G(n,m;r,s)G^{-1}(r,s;k,l)=e^{
i\pi(kl-nm)\alpha}\int_0^1d\nu\int_0^1d\nu^{\prime}e^{-i\pi\xi[(\nu+\alpha
n){\rm mod}~1]^2/\alpha}
\cdot\nonumber \\
&&e^{i\pi\xi\nu^2/\alpha}e^{2\pi i(l-m)\nu}e^{i\pi\xi[(\nu+\alpha
k){\rm mod}~1]^2/\alpha}e^{-i\pi\xi\nu^{2}/\alpha}e^{2\pi
i(k-n)\nu^{\prime}}.
\end{eqnarray}
Doing the integral over $\nu^{\prime}$ gives a factor
$\delta_{n,k}$ and we obtain
\begin{eqnarray}
&&\sum_{r,s=-\infty}^{\infty}G(n,m;r,s)G^{-1}(r,s;k,l)=e^{
i\pi(kl-nm)\alpha}\int_0^1d\nu e^{-i\pi\xi[(\nu+\alpha
n){\rm mod}~1]^2/\alpha}e^{i\pi\xi\nu^2/\alpha}
\cdot\nonumber \\
&&e^{2\pi i(l-m)\nu}e^{i\pi\xi[(\nu+\alpha
k){\rm
mod}~1]^2/\alpha}e^{-i\pi\xi\nu^{2}/\alpha}\delta_{n,k}\\
&&=e^{
i\pi(kl-nm)\alpha}\int_0^1d\nu e^{2\pi i(l-m)\nu}\delta_{n,k}.
\end{eqnarray}
Finally, doing the integral over $\nu$ we obtain a factor
$\delta_{m,l}$ and so
\begin{equation}
\sum_{r,s=-\infty}^{\infty}G(n,m;r,s)G^{-1}(r,s;k,l)=\delta_{n,k}\delta_{m,l}
\end{equation}
as we claimed. The identity in Eq.
(\ref{identities}) can be proven in the same fashion.

\pagebreak


\begin{thebibliography}{99}

\bibitem{ford}
J. Ford, G. Mantica and G.H. Ristow, Physica D 50, 493, (1991); J. Ford and
M. Ilg, Phys. Rev. A 45, 6165 (1992).

\bibitem{berry} J.H. Hannay and M.V. Berry, Physica D 1, 267 (1980).

\bibitem{keat} J.P. Keating, Ph.D. Thesis, University of Bristol, 1989.

\bibitem{ristow} G.H.Ristow, M.Sc. thesis,
Georgia Institute of Technology, 1987.

\bibitem{balazs} N.L. Balazs and A. Voros , Europhys. Lett. 4, 1089 (1987).

\bibitem{balazs2} N.L. Balazs and A. Voros, Ann. Phys. 190, 1, (1989).

\bibitem{berry2} M.V. Berry, in {\em Chaos and Quantum Physics}, ed. M.J.
Giannoni, A. Voros, J. Zinn-Justin, (North-Holland, Amsterdam, 1991).

\bibitem{cp} L.D. Landau and E.M. Lifshitz,
{\em Quantum Mechanics, Nonrelativistic Theory}, (Pergamon Press,
London, 1958); E. Merzbacher,
{\em Quantum Mechanics}, (Wiley, New York, 1961).

\bibitem{graffi} M. Degli Esposti, S. Graffi and S. Isola, ``Stochastic
Properties of the Quantum Arnold Cat in the Classical Limit'', (preprint,
University of Bologna).

\bibitem{models} Our view assumes that the torus models a mechanical
system which should be amenable to quantization. An alternative
perspective might argue that the quantum map derives from the real
physical system, which is properly represented by
a differentiable dynamic system. As such it is
the latter which should be amenable to quantization and hence not all
quantum maps need be quantizable.

\bibitem{ww} For a thorough discussion of the Wigner-Weyl representation
and its algebra see S.R. De Groot and L.G. Suttorp,
{\em Foundations of Electrodynamics}, (North-Holland, Amsterdam, 1972).

\bibitem{jaffe} For a recent application of
classical and quantum phase space dynamics see, e.g., C. Jaff\'{e} and
P. Brumer, J. Phys. Chem. 88, 4829 (1984); J. Chem. Phys. 82, 2330, (1985);
C. Jaff\'{e}, S. Kanfer and P. Brumer, Phys. Rev. Lett. 54, 80 (1985) and
references therein.

\bibitem{foot1} (a) As discussed in this section, the correct system
Hamiltonian should
reflect the system symmetry. Thus the
$\psi$ in Eq. (\ref{e0eigf}) are actually eigenfunctions of
$H=(\widehat{p{\rm mod}~b})^2/2\mu$, the
correct Hamiltonian, as distinct from $\hat{p}^2/2 \mu$.
(b) As noted in (a), this Hamiltonian was also used for the case of
wavefunctions. Thus the introduction of this symmetry adapted Hamiltonian
does not suffice to resolve the quantization problems which we are
addressing.

\bibitem{vlad} V.S. Vladimirov, {\em Generalized Functions in Mathematical
Physics}, (Mir, 1979, Moscow).

\bibitem{strangebc} Interestingly, formulating traditional quantum
mechanical problems (e.g. the rigid rotor) in the density matrix
formalism would necessitate that we
impose boundary conditions like Eq. (\ref{perbc}) on the system to obtain
the correct spectrum. We see no way of divining these boundary conditions
in the absence of wavefunction information. Hence the notion that one can
completely formulate mechanics within the von Neumann equation, as a
parallel but independent mechanics, seems to us to be difficult to
implement.

\bibitem{knabe} These operators were also introduced by S. Knabe,
J. Phys. A 23, 2013 (1990) and subsequently used in \cite{graffi}.

\bibitem{commrel} (a) The usual commutation relation, like the classical
Poisson bracket relation $\{q{\rm mod}~a,p{\rm mod}b\}=1$, is satisfied
on the interior of the torus. (b) The allowed states obey the
uncertainty principle if the phase space is large enough to hold at
least one  quantum i.e. $ab\geq h$.

\bibitem{arnold} V.I. Arnold and A. Avez,
{\em Ergodic Problems of Classical Mechanics}, (Addison-Wesley, N.Y. 1989).

\bibitem{abra} M. Abramowitz and I.A. Stegun,
{\em Handbook of Mathematical Functions}, (Dover, N.Y., 1965).

\bibitem{dumont} R.S. Dumont and P. Brumer, J. Chem. Phys. 87, 6437 (1987).

\bibitem{expect} It can be simply shown that the rule for the
evaluation of the expectation of any observable $\hat{A}(MT)$ is
$\langle
\hat{A}(MT)\rangle=\int_0^bdp\int_0^adq~(\hat{A}\hat{\rho}(MT))^{(w)}(P,Q)
=\sum_{n,m}A_{n,m}\rho_{-n,-m}(MT)$.
This is the same as the rule for evaluating averages classically.

\bibitem{mkick} A second class of mappings with $\phi=\left( \begin{array}{cc}
1+\eta\xi&\eta \\
\xi &1
\end{array} \right)$
can be quantized by
integrating from just before the $M^{th}$ kick to just before the
$(M+1)^{st}$ kick. The resulting quantum propagator for the Fourier
coefficients is
\begin{eqnarray}
&&G(n,m;k,l)=e^{
i\pi\alpha(kl-nm)}\int_0^1d\nu\int_0^1d\nu^{\prime}e^{-i\pi\xi[(\nu+\alpha
k){\rm mod}~1]^2/\alpha}e^{i\pi\xi\nu^2/\alpha}\cdot\nonumber \\
&&e^{2\pi i(l-m)\nu}
e^{-i\pi\eta[(\nu^{\prime}+\alpha
m){\rm mod}~1]^2/\alpha}e^{i\pi\eta\nu^{\prime 2}/\alpha}e^{2\pi
i(k-n)\nu^{\prime}}\nonumber.
\end{eqnarray}

\end{thebibliography}
\end{document}